# Collective decision efficiency and optimal voting mechanisms: A comprehensive overview for multi-classifier models

Harris V. Georgiou (MSc, PhD)

Dept. of Informatics & Telecommunications,
National Kapodistrian Univ. of Athens (NKUA/UoA)

Email: harris@xgeorgio.info -- URL: http://xgeorgio.info

Last Updated: 8 February 2015

**Abstract**

A new game-theoretic approach for combining multiple classifiers is proposed. A short introduction in basic Game Theory and coalitions illustrate the way any collective decision scheme can be viewed as a competitive game of coalitions that are formed naturally when players state their preferences. The winning conditions and the voting power of each player are studied under the scope of Banzhaf and Shapley numbers, as well and the collective competence of the group in terms of correct collective decision. Coalitions and power indices are presented in relation to the Condorcet criterion of optimality in voting systems, and weighted Borda count models are asserted as a way to implement them in practice. A special case of coalition games, the weighted majority games (WMG) are presented as a restricted realization in dichotomy choice situations. As a result, the weighted majority rules (WMR), an extended version of the simple majority rules, are asserted as the theoretically optimal and complete solution to this type of coalition gaming. Subsequently, a generalized version of WMRs is suggested as the means to design a voting system that is optimal in the sense of both the correct classification criterion and the Condorcet efficiency criterion. In the scope of Pattern Recognition, a generalized risk-based approach is proposed as the framework upon which any classifier combination scheme can be applied. A new fully adaptive version of WMRs is proposed as a statistically invariant way of adjusting the design process of the optimal WMR to the arbitrary non-symmetrical properties of the underlying feature space. SVM theory is associated with properties and conclusions that emerge from the game-theoretic approach of the classification in general, while the theoretical and practical implications of employing SVM experts in WMR combination schemes are briefly discussed. Finally, a summary of the most important issues for further research is presented. This report is a compact introduction to the theoretical material upon which a new expert fusion model can be designed.

# Part I – Theoretical Background

**1.1 Preface**

The ultimate goal of any pattern recognition system is to achieve the best possible classification performance for the specific problem at hand. This objective has led to the development of sophisticated algorithms and classification models in a way that best captures and enhances the underlying structure of the input space. As the complexity and intercorrelation between the classes increases, more robust and efficient classifiers have to be employed in order to provide adequate adaptability and generalization. However, the sensitivity and specificity of each classifier model can prove efficient in one case and inefficient in another, thus there is no clear indication of one single classifier design that can be considered as a universal pattern recognition solver.

Recent studies have focused in the possibility of taking advantage of this complementary performance of various classifier designs, in order to produce combination schemes for optimal fusion of multiple classifiers. Specifically, each classifier is considered as a trained expert that participates along with others in a "committee", which produces a collective decision according to some well-specified rule. The most common combination rules include the min rule, the max rule, the median rule, the majority voting rule, the averaging rule, etc.

It has been proven that all these combination rules are special realizations of two basic combinations schemes, namely the product rule and the sum rule [01]. The product rule essentially combines the classifiers' estimations on a-posteriori probabilities in a way that is consistent with the classic probability theory on independent events. When the classifiers are considered independent, i.e., when the decision of each classifier is not affected in any way by the corresponding decision made by all the other classifiers, then the join probability of the combined result can be calculated as a product of all the individual probabilities. Although the product rule is based on solid theoretical background, specifically the Bayes theory, it has provided only moderate results in practice in pattern recognition problems. One reason is that, in practice, well-trained classifiers tend to produce similar predictions, therefore the hypothesis on independent decisions is not established, although the classifiers produce their estimations separately. The other reason for poor performance is the fact that, when the combined output of all the experts is based on a product of their individual estimations, one single severe error or poor training on one expert could drive the whole group into similarly poor performance.

In contrast, the sum rule utilizes the individual experts' estimation in some additive form. All the popular combination rules, including the *majority voting*, the *median* and the *weighted averaging* rules, can be derived from the generic sum rule. It has been proven both theoretically and experimentally that these sum-based rules outperform the product rule in terms of minimum classification error (MCE) and error sensitivity. The reason for this enhanced performance in comparison to the product rule is the fact that the sum form of these rules effectively nullifies any single outliner estimations that may be produced by a poor classifier. This means that any single severe error has only minimal effect on the final combined estimation. Therefore, any combination scheme that utilizes a sum-based rule when calculating the experts' collective decision is relatively resilient to individual expert errors and this property of resiliency is increased as the number of combined experts increases.

The exact theoretical analysis of any of these combination schemes has proven noticeably cumbersome and complex over the years. Recent studies have established some general properties on total error magnitude and sensitivity, but only under strict preconditions on number of classes and their distributions [02]. For *K* classes, current rules cannot guarantee an increase of overall performance for the combined decision when each expert exhibits accuracy only at $p=1/K$. On the contrary, averaging rules usually establish increase in performance when each expert exhibits accuracy at least $p>0.5$. Much work has been conducted experimentally for identifying whether the most important factor for the final performance of such a combination scheme depends more on the exact form of the combination rule or the diversity of the experts themselves. Ensemble design methods, such as ADAboost, Random Forests and Arcing [12], focus exactly on enhancing the experts' diversity, either by means of amplifying the classification training process in areas close to the decision boundary, or by employing dataset splitting algorithms for improving the independency criterion. However, conceptualizing and quantifying diversity between classifier outputs is very challenging and it is usually based on experimental results, rather than a solid theoretical basis. As a result, most combination schemes employ optimization heuristics when designing the exact form and parameters of such a combination rule, e.g. calculating the optimal weights in a weighted averaging model [04].

Recently, more generic approaches have been proposed for designing the combination stage of such multi-classifier methods. Specifically, instead of employing a fixed rule for combining the individual outputs from the experts, a new fully functional expert node is introduced in the form of a meta-classifier: using the outputs of all the previous *K* classifiers as input, it produces an arbitrary mapping between the *K*-dim individual decision space into the final output space. This meta-classifier can essentially be any linear or non-linear model that is usually trained in the same way any of the other *K* classifiers is trained on the base data. Experimental results for models using neural networks and meta-classification nodes have been proven very efficient in many practical problems, justifying the practical gain in introducing classifier combination schemes in cases where the complexity of a pattern recognition problem requires the use of multiple, highly specialized experts [03].

Despite the fact that many practical solutions have been proposed and tested for combining multiple classifiers, the core issues of the combination problem still remains:

  (a) Is there a generic and simple way to describe current combination rules?

  (b) Is there an optimal realization of this rule for combining multiple classifiers?

  (c) If such an optimal rule exists, does it cover the non-linear combination schemes too?

Before these questions can be answered, a brief introduction on special aspects on collective decision theory has to be made first.

**1.2  Rank and response combination from classifier confidence transformation**

Any classifier combination scheme is restricted by the type and form of the individual participating classification models, as their outputs must be compatible and suited for using them in the selected combination rule. Classifiers can be generally categorized into three types, according to their output [04]:

- Type-I : the classifier produces a simple statement on class selection
- Type-II : the classifier produces a ranked list of decreasing preferability of each classes
- Type-III : the classifier produces assignments of estimated probabilities for each class

Most combination schemes use Type-I or Type-III classifiers and most of them do not allow mixed types. Nevertheless, it is possible to convert between these types in some cases, e.g. assigning simple class selections (Type-I) for the maximum estimated probability (Type-III) or for the first preference in a ranked list (Type-II). In practice, Type-II classifiers are more generic than Type-I and Type-III classifiers are more generic than Type-II, however it is possible to calculate posterior information about the estimated probabilities for each class in Type-II classifiers by examining the produced confusion matrix. In fact, many combination schemes that exploit the confusion matrix data have the advantage of using already trained classifiers with known performance, instead of applying complete re-training of all the K experts when optimizing the combination node [05].

The confusion matrix can be considered as an experimental estimation of each classifier's performance on the specific task at hand and it can be used as a measure of confidence related to the predictions produced by this particular classifier. Specifically, the elements of the confusion matrix can be used directly in a transformation model, where individual correct or incorrect counts can be rescaled and mapped in a way that is consistent with a predetermined probabilistic model. Usually, such a formulation includes a scaling or normalization function and an activation function that is consistent with a specific type of confidence measure. The log-likelihood (linear), the likelihood (exponential) and sigmoid formulations have been proposed among others as candidate functions for confidence transformation [04]. Essentially, this process ensures that the outputs of the individual classifiers are consistent and compatible with each other and with the combination rule.

When using classifiers of Type-II, i.e., classification outputs that include a ranked list of preferences to the available classes, simple counts in the form of a confusion matrix are not possible. Instead, the ranked lists have to be converted into a simple metric that defines the overall "preferability" of each class, according to multiple classification outputs. In other words, instead of counting the number of times each class is selected as the best candidate in a simple class prediction (Type-I), a measure of "desirability" is calculated by assigning desirability points or "ranks" in every sorted list of preferences and then summing them for each class separately. This scheme is known as the *Borda count method* of combining ranked lists of predictions, i.e., the outputs of Type-II classifiers. Usually, the Borda count is calculated using ranks equivalent to the position of each class index in a sorted list, which means that in a *N*-class problem the first rank position receives *N*-1 points, the second receives *N*-2 points and the last position receives 0. It is possible to allow a weighted scheme when assigning these ranking points, so that the distribution is not uniform but arbitrary. In this case, the scheme is called *weighted Borda count* or *wBorda*.

The Borda and the more generic wBorda count methods are based on the presumption that the class selection at the rank position (*i*+1) is the second most probable candidate when classifying at class selection of rank (*i*). However, due to the absence of explicit probability estimations, it is not possible to directly extract "closure" measurements between these two choices. The weights in such a combination scheme can be designed in a way that produces this closure measurement in an optimized way.

One obvious question is whether ranked list combination produces better results that simple majority selections. Indeed, in many cases the simple class selection rule and the corresponding majority selection for the final output produces different results than the one produced by Borda count [05]. However, in the case of weighted Borda count, if weights are assigned in a specific way that is consistent with a required criterion, i.e., the majority rule, then the same combination result can be achieved. In the case of the majority rule, the weights that need to be assigned are *w*=1 for the first rank position and *w*=0 for the rest of the ranks. In general, a weighting scheme can be applied in the rank positions in a way that satisfies all the requirements needed by the typical combination rules, like the sum rule and its specialized versions (min, max, median average, majority, trimmed means, spread combiner). The product rule can also be applied in the same way [05].

It has been proposed that the weights of the rank items can be variable and proportional to the measured properties of each classifier performance. When probability estimations are available (Type-III), like when using a neural network classifier, they can be used instead. In the more restricted case of simple class selections (Type-I), it has been suggested that the confusion matrix itself can be used as the basis for calculating ranked classifications, translating the a posteriori probabilities for each class into ranked list of preferences [05]. In this case, the first assumption is that the behavior of each classifier is known it is characterized by its confusion matrix, and second that this prior behavior is representative of its future behavior. The second assumption is also applied for Type-II or Type-III classifiers, i.e., even when the confusion matrix is of no importance and its validity is increased proportionally with the size of the datasets on which the classifiers are tested.

But why restrict the classifier combination scheme into a Type-II, i.e., ranked classifications? It is true that the lack of explicit probability estimation from the classifier itself produces an inherent lack of information about the classification itself. However, many widely used classification models, including Support Vector Machines (SVM), are inherently designed to produce simple class selections as output. Furthermore, the use of simple class selections or ranked lists is required when a specific weighting profile has to be applied uniformly throughout an entire "committee" of experts, like in voting schemes. In other words, experts are weighted according to their competence in an adaptive way, but the calculation of these weights is not a subject each classifier's own performance. Instead, these weights are the realization of a collective decision rule, like in a wBorda scheme. This issue and its implications will be discussed later on under the scope on weighted majority rules (WMR).

### 1.3 Cascaded versus joint parameter optimization in combination schemes

There are three general groups of combination rules that can be applied when creating a mixture of experts: (i) the fixed rules, (ii) the trained rules, and (iii) weighted combinations of confidence transformation. The fixed rules group contains all the typical rules discussed thus far, including the product rule, the sum rule and its specialized version, etc. Trained rules refer to the case of meta-classifier nodes, where an arbitrary expert is trained experimentally upon the best way to combine the outputs of $K$ experts against a given training dataset. Finally, the third group refers to models that are based on weighted order statistics, where a weight is assigned for each rank of confidence measure, rather than each classifier as in the case of weighted combination of classifier outputs. An example of this third group of combination rules is the Borda and, more specifically, the wBorda count models.

As mentioned earlier, the parameters of the combination rule itself, i.e., the weights in a weighted average or in a wBorda count, are a subject of optimization against a specific criterion, normally the minimum classification error (MCE). Similarly, for trained rules, the meta-classifier is trained according to the same optimality criterion. Since this combined classification process can be realized as either two separate stages in a cascaded model or a unified modular architecture, the optimization process can address each stage separately or jointly together. The latter case is often used for trained rules, as the meta-classifier rule can be trained jointly together with the $K$ classifiers of the first stage.

For linear trained rules, the optimal weights correspond to the relative confidence attributed to each classifier, as in the fixed rules. However, an optimization process determines the best weighting profile based on a specific training dataset, instead of using a pre-defined weighting profile as the fixed rules suggest. For rules based on typical linear discriminant functions, the optimization process can be realized as parameter estimation via regression or by applying any other formulation of typical linear classifier model. In the case of rules based on *weighted order statistics*, like the wBorda count, the most commonly used method is also regression or some other linear optimization approach.

It is worth noticing that, as mentioned in the previous section, there is an inherent relationship between applying a weighting profile to the classifiers themselves and applying a weighting profile to the rank items that they produce as output. Classifiers with high accuracy rates, i.e., a high confidence value and a proportionally high weighting factor, will produce rank classifications where the first (top) item, i.e., the one with the largest rank weight, is usually correct. This evident correlation will be explained more clearly later on, within the context of "winning coalitions" and their realization in weighted majority games (WMG).

While joint optimization of parameters should be able to provide more generic solutions, it is not yet clear if the joint approach produces better results than the cascaded scheme. In fact, recent studies have shown that the joint optimization does not improve the combination accuracy of the validation data as compared to the two-stage strategy [04]. In most cases, a simple weighted averaging rule upon trained classifiers produces the best results, even in problems of high complexity. This is attributed to the fact that, regardless of the complexity of the initial input space, the classifiers transform it to a highly restricted subspace with dimensionality equal to the number of classes available. Therefore, it is evident why optimized linear solutions, like the WMR model, may be the answer to this problem, especially when robust classifiers like SVM are considered.

**1.4 Elements of Game Theory**

In principle, the mathematical theory of games and gaming was first developed as a model for situations of conflict. Since the early 1940's, the work of John Von Neumann and Oskar Morgenstern has provided a solid foundation for the most simple types of games, as well as analytical forms for their solutions, with many applications to Economics, Operations Research and Logistics. The "zero-sum" games are able to model situations of conflict between two or more "players", where one's gain is the other's loss and vice versa. Furthermore, if all players are full informed about their opponents' decisions the game is called of "perfect information". Such games are all board games like chess and it has been proven that there is at least one optimal plan of decisions or "strategy" for each player, as well as a "solution" to the game that comes naturally as a result of all players following their optimal strategies. At the game's solution, each player can guarantee that the maximum gain an opponent can gain is kept under a specific minimal limit, defined only by this player's own strategy. Von Neumann and Morgenstern proved this assertion as a theorem called "Minimax" and in the simple case of two opposing players the solution of the game can be calculated analytically as a solution of a 2x2 set of linear equations. The consequences of the Minimax theorem have been thoroughly studied for many years after its proof. As an example, it mathematically proves the assertion that all board games, including the most complex ones like chess, have at least one solution, optimal for both players that can be analytically calculated, at least in theory.

Although the Minimax theorem provided a solid base for solving many types of games, it is only applicable in practice for the zero-sum type of games. In reality, it is common that in a conflict not all players receive their opponents' looses as their own gain and vice versa. In other words, it is very common a specific combination of decisions between the players to result in a certain amount of "loss" to one and a corresponding "gain", not of equal magnitude, to another. In this case, the game is called "nonzero-sum" and it requires a new set of rules for estimating optimal strategies and solutions. As each player's gains and losses are not directly related to the opponents', the optimal solution is only based on the assertion that it should be the one that ensures that the player has "no regrets" when choosing between possible decision options. This essentially means that, since each player is now interested in his/her own gains and losses, the optimal solution should focus on maximizing each player's own "expectations" [11]. The Minimax property can still be applied in principle when the single most "secure" option must be identified, but the solution of the game has now a new meaning.

During the early 1950's, John Nash has focused primarily on the problem of finding a set of "equilibrium points" in nonzero-sum games, where the players eventually settle after a series of competitive rounds of the game. In strict mathematical terms, these equilibrium points would not the same in essence with the Minimax solutions, as they would come as a result of the players' competitive behavior and not as an algebraic solution of the games' mathematical formulation. In 1957 Nash has successfully proved that indeed such equilibrium points exist in all nonzero-sum games[*], in a way that is analogous to the Minimax theorem assertion. However, although the Nash theorem ensures that at least one such "Nash equilibrium" exists in all nonzero-sum games, there is no clear indication on how the game's solution can be analytically calculated at this point. In other words, although a solution is known to exist, there is no closed form for nonzero-sum games until today.

It should be noted that players participating in a nonzero-sum game may or may not have the same options available as alternative course of action, or the same set of options may lead to different payoffs between the players. When players are fully exchangeable and their ordering in the game makes not difference to the game setup and its solutions, the game is called "symmetrical". Otherwise, if exchanging players' position does not yield a proportional exchange in their payoffs, then the game is called "asymmetrical". Naturally, symmetrical games lead to Nash equilibrium points that appear in pairs, as an exchange between players creates its symmetrical counterpart.

But the Nash equilibrium points are not always the globally optimal option for the players. In fact, the Nash equilibrium is optimal only when players are strictly competitive, i.e., when there is no chance for a mutually agreed solution that benefits them more. These strictly competitive forms of games are called "non-cooperative games". The alternative option, the one that allows communication and prior arrangements between the players, is called a "cooperative game" and it is generally a much more complicated form of nonzero-sum gaming. Naturally, there is no option of having cooperative zero-sum games, since the game structure itself prohibits any other settlement between the players other than the Minimax solution.

The problem of cooperative or possibly cooperative gaming is the most common form of conflict in real life situations. Since nonzero-sum games have at least one equilibrium point when studied under the strictly competitive form, Nash has extensively studied the cooperative option as an extension to it. However, the possibility of finding and mutually adopting a solution that is better for both players than the one suggested by the Nash equilibrium, essentially involves a set of behavioral rules regarding the players' stance and "mental" state, rather than strict optimality procedures [11]. Nash named this process as "bargain" between the players, trying to mutually agree on one solution between multiple candidates within a "bargaining set". In practice, each player should enter a bargaining procedure if there is a chance that a cooperative solution exists and it provides at least the same gain as the best strictly competitive solution, i.e., the best Nash equilibrium. In this case, if such a solution is agreed between the players, it is called "bargaining solution" of the game.

---

[*] Seminal works by C. Daskalakis & Ch. Papadimitriou in 2006-2007 and on have proved that, while Nash equilibria exist, they may be unattainable and/or practically impossible to calculate due to the inherent algorithmic complexity of this problem; see e.g. "The Complexity of Computing a Nash Equilibrium", 38th ACM Symposium on Theory of Computing, STOC 2006.

As mentioned earlier, each player acts upon the property of no regrets, i.e., follow the decisions that maximize their own expectations. Nevertheless, the game setup itself provides means of improving the final gain in an agreed solution. In some cases, the bargaining process may involve the option of "threats", that is a player may express the intention to follow a strategy that is particularly costly for the opponent. Of course, the opponent can do the same, focusing on a similar "threat". This procedure is still a cooperative bargaining process, with the threshold of expectations raised for both players. The result of such a process may be a mutually "deterring" solution, which in this case is called a "threating solution". Nash has formulated all these bargaining situations into a set of relatively logical axioms, under which a solution (equilibrium) exists. As in the general case of non-cooperative games, Nash's "bargaining theorem" does not provide analytical means of finding such solutions.

The notion of "bargaining sets" and "threat equilibrium" is often extended in special forms of games that include iterative or recursive steps in gaming, either in the form of multi-step analysis (metagames) or focusing on the transitional aspects of the game (differential games). Modern research is focused on methods that introduce probabilistic models into games of multiple realizations and/or multiple stages [11].

**1.5 Coalitions, Stable Sets and Indices of Power**

Nash's work on the "Nash equilibrium" and "bargaining theorem" provides the necessary means to study n-person non-cooperative and cooperative games under a unifying point of view. Specifically, a nonzero-sum game can be realized as a strictly competitive or a possibly cooperative form, according to the game's rules and restrictions. Therefore, the cooperative option can be viewed as a generalization to the strictly competitive mode of gaming.

When players are allowed to cooperate in order to agree on a mutually beneficial solution of game, they essentially choose one strategy over the others and bargain this option with all the others in order to come to an agreement. For symmetrical games, this is like each player chooses to join a group of other players with similar preference over their initial choice. Each of these groups is called a "coalition" and it constitutes the basic module in this new type of gaming: the members of each coalition act as cooperative players joined together and at the same time each coalition competes over the others in order to impose its own position and become the "winning coalition". This setup is very common when modeling voting schemes, where the group that captures the relative majority of the votes becomes the winner.

Coalition Theory is closely related to the classical Game Theory, especially the cooperating gaming [11]. In essence, each player still tries to maximize its own expectations, not individually any more but instead as part of a greater opposing term. Therefore, the individual gains and capabilities of each player is now considered in close relation to the coalition this player belongs, as well as how its individual decision to join or leave a coalition affects this coalition's winning position. As in classic nonzero-sum games, the notion of equilibrium points and solutions is considered under the scope of dominating or not in the game at hand. Furthermore, the theoretical implications of having competing coalitions of cooperative players is purely combinatorial in nature, thus making its analysis very complex and cumbersome. There are also special cases of collective decision schemes where a single player is allowed to "abstain" completely from the voting procedure, or prohibit a contrary outcome of the group via a "veto" option.

In order to study the properties of a single player participating in a game of coalitions, it is necessary to analyze the wining conditions of each coalition. Usually each player is assigned a fixed value of "importance" or "weight" when participating in this type of games and each coalition's power is measured as a sum over the individual weights of all players participating in this coalition. The coalition that ends up with the highest value of power is the winning coalition. Therefore, it is clear that, while each player's power is related to its individual weight, this relation is not directly mapped on how the participation in any arbitrary coalition may affect this coalition winning position. As this process stands true for all possible coalitions that can be formed, this competitive type of "claiming" over the available players by each coalition suggests that there are indeed configurations that marginally favor the one or the other coalition, i.e., a set of "solutions". The notion of solution in coalition games is somewhat different from the one suggested for typical nonzero-sum games, as it identifies minimal settings for coalitions that dominate all the others. In other words, they do not identify points of maximal gain for a player or even a coalition, but equilibrium points that determine which of the forming coalitions is the winning one. This type of "solutions" in coalition games is defined in close relation to "domination" and "stability" of such points and they are often referred to as "the Core". Von Neumann and Morgenstern have defined a somewhat more relaxed definition of such conditions and the corresponding solutions are called "stable sets" [11]. It should be noted that, in contrast to Nash's theorems and the Minimax assertion of solutions, there is generally no guarantee that solutions in the context of the Core and stable sets need to exist in an arbitrary coalition game.

The notion of the Core and stable sets in coalition gaming is of vital importance when trying to identify the winning conditions and the relative power of each individual player in affecting the outcome of the game. The observation that a player's weight in a weighted system may not intuitively correspond to its voting "power" goes back at least to Shapley and Shubik (1954). For example, a specific weight distribution to the players may make them relatively equivalent in terms of voting power or, while only a slight variation of the weights may render some of them completely irrelevant on determining the winning coalition [06]. Shapley and Shubik (1954) and later Banzhaf and Coleman (1965, 1971) suggested a set of well-defined equations for calculating the relative power of each player, as well as each forming coalitions as a whole [11]. The "Shapley index of power" is based on the calculation of the actual contribution of each player entering a coalition, in terms of improving the coalition's gain and winning position. Similarly, the "Banzhaf index of power" calculates how an individual player's decision to join or leave a coalition results in a winning or loosing position for this coalition, accordingly. Both indexes are basically means of translating each player's individual importance or weight within the coalition game into a quantitative measure of power in terms of determining the winner. While both indices include combinatorial realizations, the Banzhaf index is usually easier to calculate, as it is based on the sum of "shifts" on the winning condition a player can incur [07]. Furthermore, its importance in coalition games will be made clearer later on, where the Banzhaf index will come as a direct result when calculating the derivatives of a weighted majority game.

### 1.6 Collective competence and the Condorcet criterion

The transformation of cooperative *n*-person games into coalition games essentially brings the general problem closer to a voting scheme. Each player casts a vote related to its own choice or strategy, thus constituting him/her as a member of a coalition of players with similar choices. The coalition that gains more votes becomes the winner.

Condorcet (1785) was the first to address the problem of how to design and evaluate an efficient voting system, in terms of fairness among the people that participating in the voting process, as well as the optimal outcome for the winner(s). This first attempt to create a probabilistic model of a voting body is known today as the "Condorcet Jury Theorem". In essence, this theorem says that if each of the voting individuals is somewhat more likely than not to make the "better" choice between some pair of alternative options, and if each individual makes its own choice independently from all the others, then the probability that the group majority is "correct" is greater than the individual probabilities of the voters. Moreover, this probability of correct choice by the group increases as the number of independent voters increases [07]. In practice, this means that if each voter decides independently and performs marginally higher than 50%, then a group of such voters is guaranteed to perform better than each of the participating individuals. This assertion has been used in Social sciences for decades as a proof that decentralized decision making, like in a group of juries in a court, performs better than centralized expertise, i.e., a sole judge.

The Condorcet Jury Theorem and its implications have been used as one guideline for estimating the efficiency of any voting system and decision making in general. The study of effects like diversely informed voters or situations of conflicting interests have provided several aspects of possible applications in social and economical models. In the context of collective decision-making via voting schemes, the theorem provides a mean to test the "fairness" and effectiveness of such a system, as it usually constitutes the outcome that yields the best possible degree of consensus among the voting participants. Specifically, the interest is focused on how aggregate competence of the whole voting group, measured by the probability of making a correct collective decision, depends on the defining properties of the decision-making process itself, such as different coalition sizes, team setups and possible overlapping memberships. Under this context, the coalition games are studied by applying quantitative measures on "collective competence" and optimal distribution of power, e.g. tools like the Banzhaf or Shapley indices of power. The degree of consistency of such a voting scheme on establishing the pair-wise winner(s), as the Condorcet Jury Theorem indicates, is often referred to as the "Condorcet criterion". This criterion is not the only possible measure of collective competence in a voting scheme, but as it will be explained in the next section, it is very generic and it is directly linked to optimal wBorda models.

**1.7 Optimal scoring rules and Condorcet efficiency**

Let us consider a typical voting situation where an *n*-person voting group is required to cast their votes regarding a set of *M* classes, not as simple class selections but rather in the more general sense of ranking all the available options in a list of strict preference by each of the voters. Clearly there is a fixed set of possible ranking permutations and each voter essentially chooses one of them as his/her vote. The problem is to decide upon the exact combination procedure for these votes, in order to produce a result that exhibits the highest possible degree of consensus between the voters, not only in the first place as in simple majority rules, but throughout the final sorted list. One of the more widely accepted criteria for choosing the exact permutation that best reflects the cumulative will of the voting group is the Condorcet criterion [07]. As it is based exclusively on pair-wise comparisons between the voting options, i.e., every possible pair of subsequent classes in a sorted list, a system that exhibits a high degree of consistency with the Condorcet criterion should provide an aggregate ranking result that represents the best consensus solution. The degree in which such a voting scheme maximizes the consistency with the Condorcet criterion is often called "Condorcet efficiency" of the system.

It is clear that voting systems as the one described above fit the specifications of a wBorda scheme, as an optimized wBorda should also be able to produce an aggregate ranking consistent with the cumulative will of the complete voting group. Therefore, the next obvious question is whether there is a way to design such a wBorda voting scheme that maximizes the Condorcet criterion. The direct answer is that this problem is classified as NP-complete by nature, which means that due to its combinatorial nature it is not possible to be solved with algorithms of polynomial complexity. In fact, this is the reason why simple realizations of the weighting profile or even a simple majority rule is often applied in practice, in order to keep such a system simple and widely accepted by the voting group [06].

Using the notion of Condorcet efficiency of a voting system, the real problem can be focused on the exact scoring rules, often called "weighted scoring rules", than must be applied on each rank of the vote, in order to produce a result that maximizes this criterion. Two scoring systems are particularly worth noting in this context: the plurality voting, where only the top-position rank is awarded with one point ($w=1$) and all the other positions with nothing ($w=0$), and the classic Borda count, where the top-position rank receives maximum points ($w=1$), the bottom-position rank the minimum ($w=0$) and all intermediate positions a value proportional to the exact rank [08].

Under the scope of weighted scoring rules and the more general theory of weighted order statistics, it has been proven theoretically that for three classes and $n$-person voting group, the scheme that maximizes the probability that any pair-wise contest in the final ranked list will be consistent with the pairwise majority rule, is in fact the Borda count model. This means that if the system should sort the list of winners in a way that is consistent with the pairwise majority rule, then the Borda count scheme can accomplish this. Similarly, if the majority criterion is instead replaced with the more generic Condorcet criterion, a specific wBorda model with non-uniform weighting profile is the optimal solution in this case [08]. This diversity between the two optimized scoring rules comes from the fact that the Condorcet criterion suggests a stricter rule of optimality than the simple majority and this is why the existence of a Condorcet winner is not always guaranteed [05]. In [08], a geometrical realization of the wBorda design process has been suggested and results have shown that for three candidate classes the middle-position rank has to be assigned with a weight value of less than zero, i.e., different than the classic Borda rule, in order to obtain maximum Condorcet efficiency. Current theoretical results are not sufficient to support any generic statement regarding the design properties of such optimal schemes. Furthermore, the high degree of complexity prohibits the analytical theoretical study of such systems, in order to produce generic constructive methods for wBorda of maximum Condorcet efficiency.

### 1.8 Majority functions and Banzhaf numbers

Let us now focus in the case of dichotomous choice situations where there are only two candidate classes to vote for. This is clearly a simpler problem in terms of pattern recognition, since an input has to be classified in either one of the two available choices, "true" or "false", "positive" or "negative", "benign" or "malignant". As there are only two available class choices, the Borda count, which is used when class rankings are considered, reduces to the simple class selection scheme and the resulting majority rule that is used in practice.

Dichotomy choice situations have been the center of many analytical probabilistic studies within the scope of voting systems, primarily because of the simplicity of the probabilistic formulations of such models. A dichotomy choice can be easily modeled as a binomial distribution and the combined result in a *n*-player voting game becomes a product of the corresponding "skill" probabilities of the individual players. Then, a combined decision rule can be formulated according to the aggregate choice that is supported by the largest combined probability, i.e., the choice that corresponds to the maximum degree of consensus among the players. These decision rules are called "majority functions" and in the special case where all players and classes are accompanied with the same weight, the simple majority rule emerges as a natural result.

In terms of coalition games, the simple majority functions are modeled in a way which is much more trivial than the generic wBorda scheme that was presented in previously. Again, if special voting situations like abstains and veto are not allowed, the choice of either one of the two available classes automatically assigns every participating player into one of the two possible coalitions. Classes may or may not be weighted with the same "value" or "importance", while the players themselves may be accompanied with a weight or "reliability" value too. In any case, if a linear rule is applied to accumulate and combine all individual choices in order to make a final collective estimation, a weighted majority rule emerges. The threshold of the majority decision may also be altered in a way that requires not only relative majority, but a majority value higher than a specific decision threshold. In practice, this means that a bias may be used in the weighted majority function in order to ensure that the final majority outcome is valid only if it attains a specific confidence level.

The analysis of the majority functions is often restricted to the non-weighted case, as they are much easier to analyze within the scope of classic probabilistic theory. In fact, this special case of majority functions can be easily related to the Banzhaf power index [07]. Specifically, if the collective efficiency is to be calculated as a function of the individual "skill" probabilities of the players, the partial derivatives of the majority functions against these probabilities are calculated. These derivatives essentially estimate the number of "shifts" that a player with a specific skill probability can cause in the winning position of any winning coalition, i.e., it is exactly what the Banzhaf power index stands for. This assertion can be extended for the weighted majority functions as well, in a slightly more complex probabilistic form. Based in this very important conclusion, it is possible to translate many of the properties of coalition games into properties that are directly linked with each player's skill.

The first and extremely important conclusion from studying the Banzhaf numbers as the derivatives of a majority function is the fact that the maximum of these derivatives should point to the configuration where maximum Banzhaf power occurs for all the voting players. Indeed, it can be easily proven that maximum Banzhaf values correspond to individual skill probabilities close to 0.5 for all voting players. That is, the vote of each voting member reaches its maximum "value" when all players have the same average skill for making correct estimations. Interestingly enough, that is exactly what the Condorcet Jury Theorem suggests in a more generic way. An electorate system with high Banzhaf number corresponds to a high level of collective competence, which in turn is obtained for a high level of "democracy" in the sense of an equitable distribution of decisional power among the voters [07]. As high Banzhaf numbers indicate a high degree of democracy among the voting members, the decentralized option for making collective decisions is, again, asserted as the optimal way – this time in a more strict mathematical statement. This is perhaps a sufficient justification for using ensembles of independent experts with only moderate efficiency, rather than one single expert of the very high efficiency.

But what about the distribution of power within the voting group itself? Using the same formulation of Banzhaf numbers as derivatives of the corresponding majority functions, Berg [07] and Taylor and Zwicker [06] have stated some very interesting results regarding the optimal structure and distribution of voting power of such a system. Specifically, it has been proven that in any non-weighted majority function, the sub-division of the players into teams that subsequently take part in a second-stage indirect voting scheme results in a loss of individual decisional power, in terms of how a single player can affect the final result with his/her vote [07]. This essentially means that it is better to have *KxN* voters in one voting group than to split them into *K* teams of *N/K* members each for voting via *K* representatives. Boland (1989) has proved that although the Condorcet Jury Theorem also stands true for indirect voting systems, splitting a voting group into teams essentially lowers the majority threshold necessary for a coalition to become a winning one, thus reducing the probability for a collectively correct decision [07]. This implies that for the same overall voting group size and individual skill probabilities $p>0.5$, an indirect voting scheme always has less reliability than a corresponding direct system, which in turn favors combination methods with the least possible integration stages. This effect is more evident for systems that employ a relatively large number of voters, rather than small-sized systems where this difference is expected to be minimal [07].

In terms of team sizes versus number of teams, Boland (1989) has also proved that in an indirect system, a large number of small teams are collectively more effective than a small number of large teams. In the extreme case where each team includes only one voter, the indirect system becomes direct, i.e., with no representatives, which is the strictly more efficient voting structure as noted earlier. Special studies have been carried out for situations of teams with unequal number of members or for overlapping memberships. Again, as in the case of single voting players, it has been proven that the collectively more efficient choice is splitting the voters into teams of equal size and distinct memberships, i.e., in way that favors equal distribution of voting power in every case [07].

The overlapping membership case can be viewed as a situation where some of the players are allowed to participate in more than one representative team, in other words to affect the final outcome with more than one votes. This is essentially equivalent to having an increased reliability or weight assigned to these players. As mentioned earlier, the weighted majority functions are a generalized version of the ones that have been studied within this scope. As it turns out, the collective decision efficiency may benefit from such an overlapping membership, i.e., a weighted voting scheme, only when the players with multiple votes exhibit a skill probability higher than a specific threshold. This higher than the average skill level *P* of the rest of the voting players threshold and it depends on this average skill level *P* and the total number of *N* voters, but not on the number of *K* teams [07]. This conclusion favors the application of weighted versus non-weighted majority functions in theory, but it does not specify an optimal way to find out which players to favor, in other words how to calculate these weights. This issue will be addressed later on within the context of weighted majority games (WMG). Generally speaking, analyses in terms of voting games and distribution of power are not common in the literature on the Condorcet Jury Theorem. Austen-Smith and Banks (1996), as well as Berg [07], stress the importance of a game-theoretic approach to collective decision making.

### 1.9 Weighted Majority Games and Weighted Majority Rules

In most cases, majority functions that are employed in practice very simplistic when it comes to weighting distribution profile or they imply a completely uniform weight distribution. However, a specific weighting profile usually produces better results, provided that is simple enough to be applied in practice and attain a consensus in accepting it as "fair" by the voters. Taylor and Zwicker (1991) have defined a voting system as "trade robust" if an arbitrary series of trades among several winning coalitions can never simultaneously render them losing [06]. Furthermore, they prove that a voting system is trade robust if and only if it is weighted. This means that, if appropriate weights are applied, at least one winning coalition can benefit from this procedure.

As an example, institutional policies usually apply a non-uniform voting scheme when it comes to collective board decisions. This is often referred to as the "inner cabinet rule". In a hospital, senior staff members may attain increased voting power or the chairman may hold the right of a tie-breaking vote. It has been proven both in theory and in practice that such schemes are more efficient than simple majority rules or any restricted versions of them like trimmed means.

Nitzan and Paroush (1982) have studied the problem of optimal weighted majority rules (WMR) extensively and they have proved that they are indeed the optimal decision rules for a group of decision makers in dichotomous choice situations [09]. This proof was later (2001) extended by Ben-Yashar and Paroush, from dichotomous to polychotomous choice situations [13]; hence, the optimality of the WMR formulation has been proven theoretically for any *n*-label voting task.

A WMR is in fact a realization of a weighted majority game (WMG), where a group of players with arbitrary skill levels form coalitions of similar interests but different estimations. The WMGs are a well-known subgroup of coalition games where there are only two possible coalitions, each related to one of the two class options available. In this form of gaming, there is no need of class ranking schemes as in wBorda, thus the classification problem reduces to the optimal design of a combination rule between two extreme options. These optimal combination rules are similarly called WMR and the proof that they are linear in nature limits the problem to the estimation of an optimal (non-negative) weighting profile for the voters.

The weight optimization procedure has been applied experimentally in trained or other types of combination rules, but analytical solutions for the weights is not commonly used. However, Shapley and Grofman (1984) have established that an analytical solution for the weighting profile exists and it is indeed related to the individual player competencies or skill levels [09]. Specifically, if decision independency is assumed for the participating players, the optimal weights in a WMR scheme can be calculated as the *log-odds* of their respective skill probabilities, i.e., $W_i = \log(O_i) = \log(P_i/(1-P_i))$. Interestingly enough, this is exactly the solution found by analytical Bayesian-based approaches in the context of decision fusion of independent experts [12]. The optimality assertion regarding the WMR, together with an analytical solution for the optimal weighting profile, provides an extremely powerful tool for designing theoretically optimal collective decision rules. Usually, the winning coalition in a WMR is the one that accumulates the relative weighted majority, which is more than half the sum of weights. If a bias is also applied as a confidence threshold, then the simple weighted majority rule becomes the "cogent" weighted majority rule. In this case, a region of "stalemate" or "no-decision" is created and the existence of a winning coalition in a WMR is not guaranteed.

There is an equivalence relation on the WMRs, whereby two WMRs are equivalent if they produce the same decision function, i.e., the same outcome for each decision profile produced by the participating players. Therefore, from all the possible realizations of a WMR of a given dimension, it is possible to identify a closed set of unique WMRs that are able to produce all the possible combination outcomes with a normalized version of the weights [09]. As an example of such closed set of solutions, the unique WMRs for *n*=4 voting players are: S1={1,0,0,0}, S2={2,1,1,1} and S3={1,1,1,0}. The S1 and S3 solutions clearly implement the restricted majority rules with odd number of voters, while S2 implements a simple majority rule with a tie-breaking option for one of the players. It can be proven that all other realizations of 4-player WMRs can be mapped into one of these three unique WMRs.

The calculation of such a set of WMRs is cumbersome and it is generally an NP-complete problem. However, there have been analytical studies for WMRs of dimension up to seven players. For example, in the set of all 84 WMGs between five players, only 7 of them are not transformations of others. Von Neuman and Morgenstern (1944) identified the 21 unique WMGs for six players, while Isbel (1959) and Fishburn and Gehrlien (1977) identified the 135 unique WMGs for seven players. Karotkin (1994) coded a special algorithm to identify the WMRs or the WMGs for any group size and proposed a graph-based method for illustrating the decision-based "closeness" of such WMRs in a three-dimensional space, called the "network of WMR" [09]. Using directed edges, the network of WMR can identify the node that is optimal for a given set of players' skill probabilities profile, i.e., the WMG solution that is in fact the one that the log-odds calculates for the corresponding weights. The practical use of such a graph-based representation of WMRs is that it can suggest optimal substitutions of the theoretically optimal WMR with best sub-optimal simplified realization [10]. This is extremely important in real life problems where simple collective decision rules with straightforward application are needed.

### 1.10 Weighted Majority Rules and Condorcet efficiency

The efficiency of a WMR is defined as the likelihood that it will resolve in the correct choice, given the skill probabilities of the participating voters. These likelihood functions are quite difficult to calculate in practice due to the fact that the number of possible decision profiles is a combinatorial enumeration problem. As a result, it is also difficult to compare relative efficiencies between different WMRs. However, since a WMR is proved to be the optimal structure in the sense of collective decision competence, the corresponding weighting profile that is optimal for a given set of skill probabilities of the participating voters should be the actual realization of the theoretically most efficient voting scheme.

As mentioned earlier, in the case of multiple class options where class rankings are necessary, it is possible to find a wBorda scheme that maximizes the Condorcet efficiency of such a voting system, although this problem is generally NP-complete. When this setup is reduced into the dichotomous choice situation where there are only two classes available, this model becomes the theoretically optimal formulation of the WMRs. However, in this case there is an analytical solution for the optimal weighting profile that is not NP-complete, although the complete enumeration of all the unique WMRs that can be implemented in practice is of that complexity. As a result, the next question is at what degree a WMR can be viewed as an optimal solution to a WMG in the Condorcet sense.

To answer this question, the notion of "bias" or "confidence threshold" in a WMR has to be reviewed under a new perspective. In wBorda schemes, each class ranking position is scored with a specific weight and the corresponding scoring rule is considered optimal in the Condorcet sense if it maximizes the Condorcet criterion. Similarly, in the case of two class problem, the simple choice between the one or the other choice essentially implies a similar preference ranking regarding the classes. Therefore, both the first (proposed) class choice and the second (rejected) class choice can be assigned with a scoring value, i.e., a weight, that can be incorporated into the standard WMR formulation. These scoring values are not a subject of the players' skill probabilities, since the efficiency of each player affects only the corresponding weight it receives within the WMR function, not the scoring result of selecting or rejecting a class. In a sense, the scoring of class selection or rejection adds a weighting scheme in the second dimension of the WMRs, that of the classes.

Using this new more generalized formulation of WMRs, it can be easily proven through linear transformations that this class scoring essentially produces a "positive" or "negative" bias to the accumulated result of the standard WMR. Therefore, a decision threshold can be shifted towards the one or the other class accordingly, based not only on which exactly of the players selected it but also the mere (weighted) count of the times it was proposed or rejected by all the players. If all players' votes are weighted exactly the same, then this new scheme is a two-class realization of a wBorda count model. But the wBorda model has been already proven as adequate for providing an optimal voting realization in the Condorcet sense. In fact, adding a bias to both classes according to their selection count is mathematically equivalent to setting the WMR decision threshold at a value other than half the sum of the weights, i.e., "biased" towards one of the classes. Not surprisingly, this new generalized version of the WMR can also be considered adequate for implementing voting schemes that maximize the Condorcet criterion, i.e., the exhaustive pair-wise ranking contest between the coalitions.

The assertion that WMRs are optimal realization of combinations schemes in dichotomy choice situations has some extremely significant implications in the way the WMRs can be used as a unified template model for creating optimal collective decision systems. These linear formulations of WMGs are optimal in the MCE sense but additionally they can be designed to be optimal in the Condorcet sense. Dichotomy choice situations are simple enough so that a Condorcet winner, that is the overall top-ranked class, is also the majority winner, which is simple the class that received the most votes. If weights are applied to the players, then the simple majority rule becomes a weighted one. If scores are also applied in the "support" or "reject" options (ranks) of the classifications, then a two-class wBorda count model can be realized in a way that maximizes the Condorcet criterion. In practice, this second case is equivalent to imposing a collective decision threshold other than half the sum of the weights. While the players' optimal weighting profile in the WMR solves the problem of how to combine their individual decisions in an MCE-optimal way, the class scoring provides the means the design the voting system in a way that is also optimal in the Condorcet sense. The conditions under which these two properties can be satisfied simultaneously remains an open issue.

# Part II – Applications to Pattern Recognition[†]

## 2.1 Basic framework

In part-I of this report it was suggested that classifiers providing "hard" decisions, or other types of classifiers with translated output to a set of distinct choices, can be used as the basis of a general combination procedure for providing a collective decision system. Furthermore, it was described how this model can be effectively fused into a game-theoretic approach of the combination problem that finally leads to coalition games and collective decision theory. This section describes how these models can be realized and implemented as practical systems within the scope of Pattern Recognition.

In order to combine classifier outputs in an optimized way, first it is necessary to convert their posterior accuracy probabilities into quantitative measures of evidence regarding their past performance. For classifiers of "hard" decisions, the confusion matrix is a very descriptive and perhaps the most practical way to do this. Specifically, the confusion matrix itself can be translated into class rankings and conditional probabilities estimations, as it was suggested earlier. Furthermore, the use of the confusion matrix is completely compatible with any extension that introduces the notion of "risk" into the classification process.

Decision-critical applications, like in medical diagnostics, require strict distinction between the various cases of correct and incorrect predictions. This means that a specific weight is assigned for every such classification case, in the form of a positive "gain" for correct predictions or a negative "loss" for misclassifications. When combined with the corresponding posteriori probabilities of the classifier, it is possible to calculate the expected statistical "risk", i.e., the average gain or loss that this particular classifier can produce. If the classifier is trained from the start by applying optimization criteria based on risk factors, rather than simply the classification accuracy, then the process is a "risk-based" rather than "error-based" training of the classifier. Not all classifier architectures are fit to be implemented as risk-based models, primarily due to the fact that the introduction of risk factors within the feedback process of the training may result in severe instability and failure. However, the notion of risk embodies a much more generalized viewpoint of the classification problem and it is very important in real-world applications.

Using risk-based models for the classifiers, the game-theoretic approach of collective decision systems becomes much more comprehensible. The efficiency of each participating "player" is now measured not simply in a sense of absolute accuracy but in the scope of average "gain" in each run of the game. Therefore, every combination scheme also embodies the same notion of maximizing the collective "gain" or, equivalently, minimizing the collective "loss", by employing an optimal combination rule. This risk-based approach is also valid for a coalition's winning stance against the others, as well as the expected payoff from the whole game, since a winning coalition's gain coincides with the overall gain

---

[†] Comments in this section are subject of own study and experimental verification, conducted during the author's PhD work, 2001-2008 and on. Almost all of the proposed items have been addressed, experimentally tested and subsequently published in various conference, journal and open-access papers. For detailed description of the theoretical and practical aspects of applying these ideas in the context of novel classifier combination architectures, see e.g. [14-16]:

- "A Game-Theoretic Approach to Weighted Majority Voting for Combining SVM Classifiers", Harris Georgiou, Michael Mavroforakis, Sergios Theodoridis. Int. Conf. on ANN (ICANN), 10-13 September 2006 @ Athens, Greece. Ref: S.Kollias et al. (Eds): ICANN 2006, Part I, LNCS 4131, pp. 284-292, 2006.
- "A game-theoretic framework for classifier ensembles using weighted majority voting with local accuracy estimates", H. Georgiou, M. Mavroforakis, arXiv.org preprint (en)(arXiv:1302.0540v1 [cs.LG]).
- "Algorithms for Image Analysis and Combination of Pattern Classifiers with Application to Medical Diagnosis", H. Georgiou, PhD thesis summary (en)(arXiv:0910.3348v1 [cs.CV]).

of the collective decision rule. Therefore, the formulation of this type of gaming under the scope of WMGs and the corresponding WMRs comes very naturally.

Since the WMR have been proven as the optimal combination rules in dichotomy choice situations, it is very interesting to examine the conditions under which these optimality assertions stand true for trained classifiers in the place of players. It is expected that, as these classifiers are more or less dependent with each other due to similar architectures or training datasets, the design of optimal WMRs for combining them can not be realized in completely closed form. Instead, the calculation of the exact weighting profile requires the exploitation of various statistical and structural properties of the feature space, as well as the correlation of input patterns and classes. Most ensemble techniques exploit these properties by enhancing classification regions of special interest, like points close to the decision boundary. This issue has been noted earlier within the scope of voting systems, specifically in relation to the diversity between the experts. In practice, many ensemble methods that employ maximization of diversity essentially increase the degree of independency between the participants. Since an increased level of independency provides the means for a collectively efficient decision, it is not surprising to see that the results from the Pattern Recognition viewpoint coincide with the ones inferenced by the game-theoretic approach, where Banzhaf and Shapley indices of power can be considered as measures of diversity among the participants.

**2.2 Adaptive realizations of WMRs**

As it was mentioned previously, assumptions of complete independency between the classes and their corresponding coalitions in WMGs, as well as between the experts are never completely true. Thus, it is necessary to enhance the combination process in a sense that takes into account these types of correlations. Tresp and Taniguchi (1996) have suggested that a combination scheme which uses a fixed or a weighted majority rule should exploit the properties of the statistical distributions of the classes at hand. Using a Gaussian approximation, they have shown how the efficiency of such a combination rule can be improved if the mean and variance of each class are used when calculating the parameters (weights) of the combination rule. However, the standard WMR approach for optimal combination of experts does not include such an adaptive scheme. Furthermore, the variance-based weighting method of Tresp and Taniguchi impose further assumptions and restrictions to the distributions of the classes, which can be non-Gaussian in general.

Instead of employing a fixed statistical approximation for the complete class distribution, a new fully adaptive approach can be designed on a lower level. Specifically, since the distribution of each class, i.e., the topological couplings between the individual training samples of the class, affects the exact weighting profile of the classifiers that is optimal in some error-based or risk-based criterion, then a topological measure of "closure" between any arbitrary pair of samples should be used instead of the statistical approximation of their distribution. In other words, instead of checking how well an unclassified sample fits the statistical distribution of the one or the other class, it should be checked under a criterion that measures how close it is with the identified members of each class, preferably with the most representative ones. As in the case of the variance-based method, this measure should be used as a quantitative guideline regarding the degree of "responsibility" that each class manifests over this particular point in the feature space. In terms of classification, it is a statistical method of measuring how much a class is accountable for this new sample, but in a more invariant way than that of using Gaussian approximations for the class' distributions.

When this fully adaptive version is adopted within the scope of the WMRs, each classifier's posterior probability or "skill", already known from the training process, can be adjusted according to how "far" or "close" an unclassified sample lies with respect to the already known members of this class. From a clearly topological point of view, this is a way to minimize the "structural risk" of the classification problem by introducing a bias or "preference" towards the class that seems to be more responsible for this region of the feature space. Although this is what each of the participating classifiers does on its own, this adaptation process essentially adjusts the process of evaluating the optimal WMR to the non-symmetrical structural properties of the feature space. Therefore, it is also expected that the weighting profile calculated for the WMR design would be optimal, not in a global but rather in a more local sense.

### 2.3  A link with the theory of SVM

SVM architectures provide the necessary foundations for a theoretical sound framework of optimal classifiers. The use of special form of kernel functions essentially makes them equivalent almost to any type of linear and non-linear pattern classification model. But the solid theoretical background of this type of classifiers makes them ideal in situations where their performance and consistency is required for studying collective decision rules.

The linear form of the WMRs makes the combination process very simple, not only in terms of calculating the final outcome of the group decision, but also in the scope of statistical properties of this decision rule in relation to each classifier's own properties. SVM theory states that the structure of the SVM classifier permits the linear transformation of a number of kernel functions into one combined linear form. Furthermore, if each of the kernel functions is well-defined under the typical constraints for SVM kernels, then their linear combination is also a well-defined SVM kernel function. This assertion is of extreme importance when viewed under the perspective of WMGs and WMRs. In essence, if all participating classifiers are assumed to be SVM realizations, then the optimal WMR itself defines a new compound SVM kernel, i.e., an SVM meta-classifier.

This conclusion is an adequate justification on why such a combination rule does not need to be more complex than a linear transformation of each expert's assessment: if every expert is of adequate skill and acts independently from the others. This means that such an expert can moderately efficient in the complete feature space or, alternatively, well-adapted to only a part of the complete feature space. In the first case, the WMR is optimal in the global sense in a way that combines the group of experts in the most promising manner, while the second case corresponds to the fully adaptive WMR realization that was proposed in the previous section.

It should be noted that, although SVM classifiers are generally design for "hard" decision classifications in dichotomy choice problems, it is not difficult to design a set of SVM classifiers that are "specialized" in one of $N>2$ classes if the problem requires it – this is essentially the one-versus-all classification mode when applying binary pattern classifiers in multi-class tasks. Furthermore, their internal structure that is based on support vectors, i.e., class members that primarily define the classification outcome, is well-suited for the design of robust topological measures of "closure" between a class and a new unclassified sample, based on distance transformations from the support vectors of each candidate class. In this sense, even a two-class problem that is solved by a single SVM classifier can be viewed as a coalition game in the form of WMG, solved by an optimal WMR with weights and bias proportional to the class' distribution characteristics, and an SVM kernel function that effectively transforms the original non-linear feature space to a linear space of higher dimension that can be solved by this WMR.

## 2.4 Conclusion and further work

The material presented briefly in part-I of this report clearly define a solid background for a game-theoretic approach to the problem of classifier combination in its general form. A set of theoretic formalizations lead to some very intuitive and simple solutions to this problem in the general sense, especially in dichotomy choice situations.

An extension of these theories to the area of Pattern Recognition can be easily inferenced. Specifically, there are three main issues of special interest:

1. The introduction of a theoretically solid model for using transformations of posterior probabilities of the classifiers, e.g. by the confusion matrices, in combination with the general framework of risk minimization, either in a post-training sense or within the training process itself (risk-based training).

2. The formulation of a complete and fully adaptive realization of the WMR model that incorporates the non-symmetrical properties of the underlying feature space, when calculating the optimal weighting profile for the combination rule.

3. The study of theoretical and practical implications of introducing SVM classifier architectures as voting players in a WMG, primarily in the scope of completeness and optimality of such a solution in the general sense.

A study that addresses all these three issues should first focus on the theoretical aspects and formal definitions of any new models and algorithms, and subsequently conduct experimental tests on well-known classification problems where comparative results are available for other typical classifier combination schemes. Based in the theoretical assessment presented in this study, it is expected that such a game-theoretic approach of collective decision, along with the application of SVM classifiers, will produce results of at least the same degree of success as the best ensemble methods available today.

**References (short)**